# An Oscillator-based MaxSAT solver

Mohammad Khairul Bashar[1], Jaykumar Vaidya[1], Antik Mallick[1], R S Surya Kanthi[1], Shamiul Alam[2], Nazmul Amin[2], Chonghan Lee[3], Feng Shi[4], Ahmedullah Aziz[2], Vijaykrishnan Narayanan[3], Nikhil Shukla[1]*

[1]Department of Electrical and Computer Engineering, University of Virginia, Charlottesville, VA- 22904, USA

[2]Department of Electrical Engineering and Computer Science, University of Tennessee, Knoxville, TN 37996, USA

[3]Department of Computer Science and Engineering, Pennsylvania State University, University Park, State College, PA 16801, USA

[4]Department of Computer Science, University of California- Los Angeles, Los Angeles, CA 90024, USA

*e-mail: ns6pf@virginia.edu

**Abstract**

The quest to solve hard combinatorial optimization problems efficiently- still a longstanding challenge for traditional digital computers- has inspired the exploration of many alternate computing models and platforms. As a case in point, oscillator networks offer a potentially promising energy efficient and scalable option. However, prior oscillator-based combinatorial optimization solvers have primarily focused on quadratic combinatorial optimization problems that consider only pairwise interaction among the oscillators. In this work, we propose a new computational model based on the maximum entropy production (MEP) principle that exploits higher order interactions among the oscillators, and demonstrate its application in solving the non-quadratic maximum satisfiability (MaxSAT) problem. We demonstrate that the solution to the MaxSAT problem can be directly mapped to the entropy production rate in the oscillator network, and subsequently, propose an area-efficient hardware implementation that leverages Compute-in-Memory (CiM) primitives. Using experiments along with analytical and circuit simulations, we elucidate the performance of the proposed approach in computing high-quality optimal / near-optimal solutions to the MaxSAT problem. Our work not only reveals how oscillators can solve non-quadratic combinatorial optimization problems such as MaxSAT but also extends the application of this dynamical system-based approach to a broader class of problems that can be easily decomposed to the MaxSAT solution.

**Introduction**

Modern information processing largely relies on digital computers. Problems are solved by creating their Boolean abstractions which are mapped onto the underlying hardware platform, conventionally, realized using CMOS-based digital switches. While this model of computing has been effective at solving many problems, there is a large class of computing problems that are still considered intractable to solve using digital computers[1,2]. Commonly referred to as NP-hard problems, many combinatorial optimization problems including the archetypal class of Boolean satisfiability problems – the focus of the present work, belong to this complexity class. Computing their solutions typically entails an exponential increase in computing resources (time, memory) as the problem sizes increase. This is due to the exponential increase in the size of the solution space that the computational platform must search in order to find the optimal solution. Consequently, this has motivated the investigation of alternate computing models[3-7] and their corresponding hardware implementations that can potentially provide a more efficient approach to solving such problems. In this work, we demonstrate an analog approach using oscillators to solve Maximum Satisfiability (MaxSAT), defined as the problem of determining the maximum number of clauses of a given Boolean formula in conjunctive normal form (CNF), that can be made true by an assignment of truth values to the variables of the formula[8]. Solving MaxSAT efficiently is of immense practical relevance to areas such as software and data analysis[9,10], automotive configuration[11], electronic design automation[12], scheduling[13,14], AI planning[15], probabilistic reasoning[16,] among others.

Combinatorial optimization problems (such as MaxSAT considered here) entail the maximization/minimization of a function within a discrete/combinatorial domain. Dynamical systems[17-26] (here, interacting oscillators) can provide a novel approach to solving such problems if the combinatorial function can be mapped to a physical property (e.g., energy, entropy, etc.) that a system will naturally aim to maximize/minimize. If such an equivalence can be established, then as the system evolves to minimize (e.g., internal energy) or maximize (e.g., entropy) the physical quantity, it also "computes" the solution to the problem. Further, it is believed that the simultaneous evolution of the oscillators through a continuum of states facilitates rich spatio-temporal dynamics that can provide a highly parallel approach to searching for the ground state in the high-dimensional solution space, potentially providing a performance benefit[27-36]. Examples of prior simulation works that leverage this basic idea (not with oscillators) to solve the MaxSAT problem include the gradient descent-based approach proposed by Molnar *et al.*[37]- which requires the consideration of exponentially growing auxiliary variables, and memcomputing – based on computation in memory[38], among others. In this work, we propose and demonstrate a compact, low power oscillator-based analog architecture to solve the MaxSAT problem.

**Results**

**Oscillator system design.** Recently, electronic oscillators (and their coupled systems) have been successfully shown to solve combinatorial optimization problems[39] such as maximum cut (MaxCut)[40-45] and graph coloring[46-48] based on quadratic optimization models such as the Ising model. These models consider pairwise ($J_{ij}\sigma_i\sigma_j$) interaction among the spins which are then mapped to the coupled oscillators as spin ≡ oscillator;

and (pairwise) interaction ≡ coupling element among (pairs of) oscillators. However, mapping problems such as Boolean satisfiability to the Ising / QUBO (quadratic unconstrained binary optimization) models can itself be computationally expensive[49,50]. Therefore, in this work, we propose a model that exploits non-quadratic interaction among the oscillators, and show its application in solving the MaxSAT problem.

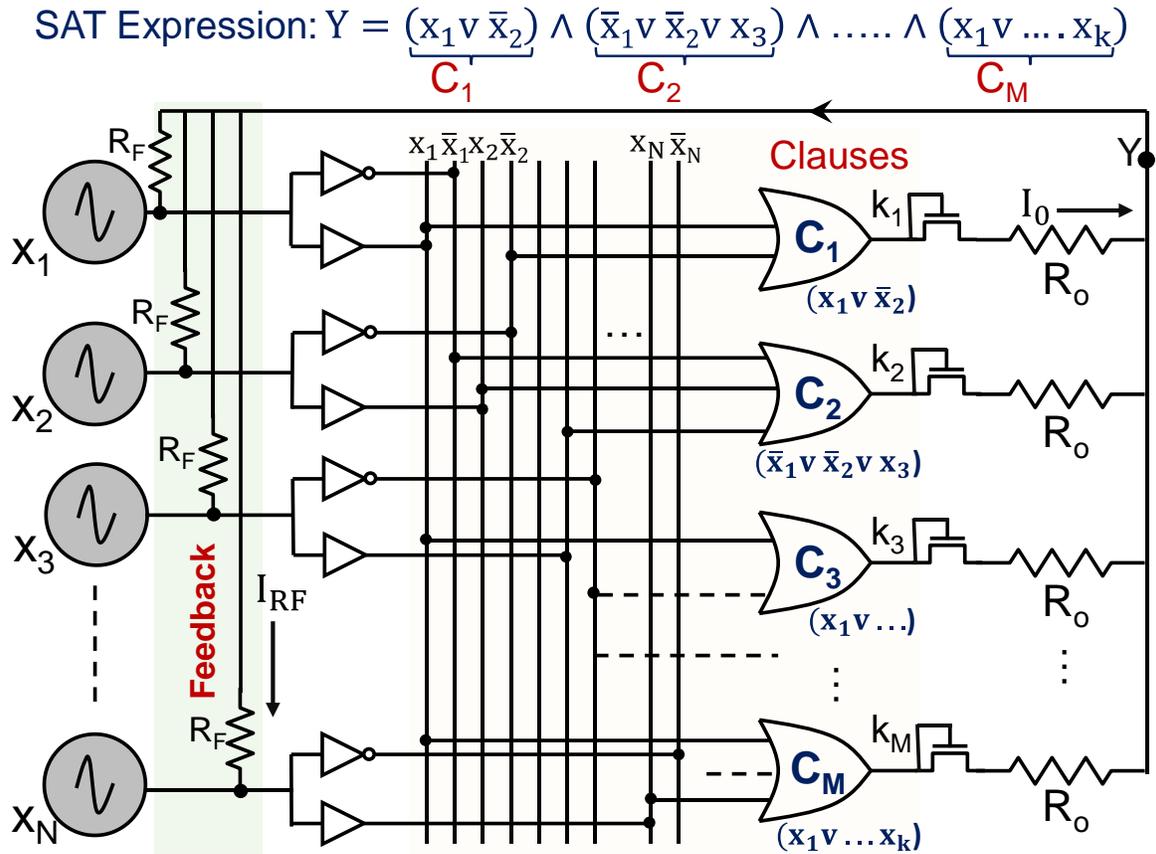

**Fig. 1| Conceptual design of oscillator system to solve MaxSAT.** Schematic of the conceptual oscillator network used to solve the MaxSAT problem. The entropy maximization in the network directly maps to solution of the MaxSAT problem. The variables of the Boolean expression are represented by the oscillators whereas the clauses are defined using the logical OR gates.

Figure 1 shows a schematic of the conceptual oscillator circuit proposed to solve the MaxSAT problem for a Boolean expression (expressed in the CNF form[51]) with N variables and M clauses. Each variable $x_j$ ($j = 1, 2 \dots N$) is represented by an oscillator, and each clause is realized using an OR gate with inputs from the oscillators corresponding to the variables appearing in the clause. If a variable appears in the negated (normal) form in a clause, an inverting (non-inverting) buffer is used to threshold the output. The output of the $i^{th}$ *(i=1, 2… M)* OR gate (i.e., $i^{th}$ clause) is represented by $k_i$ = 1 ($\equiv V_{DD}$) or 0 ($\equiv 0\ V$); $I_0$ is the output current of the gate (driving $R_o$). We model the transition in $k_i$ (low-to-high and high-to-low transitions) using a continuous-time function (the *tanh()* function is used here). A unidirectional switch (shown as a diode connected transistor) is also considered at the output of each gate to ensure that no reverse current flows when the output of the gate is zero. Subsequently, the current output of each OR gate (corresponding to the clause) is summed at the output node *Y,* and used as feedback to each oscillator. Therefore, in terms of the physical system, calculating the MaxSAT solution is equivalent to the system maximizing the number of OR gates ($\equiv$clauses) having an output of logic 1, or, in other words, maximizing $\sum_{i=1}^{M} k_i$.

To show that the above circuit will indeed aim to maximize $\sum_{i=1}^{M} k_i$, we consider the Maximum Entropy Production (MEP) principle which states that a dissipative system (subject to physical constraints imposed by laws such as Kirchhoff's current law (KCL), etc.) will maximize its total entropy production rate $\left(\frac{dS}{dt}\right)^{52}$, where $S$ is the entropy produced by the system. For the above system, the total entropy production rate can be calculated as:

$$\frac{\partial S}{\partial t} = \frac{\partial}{\partial t}\left(\sum_{i=1}^{M} k_i S_{R_o} + \sum_{j=1}^{N} S_{R_F} + \sum_{j=1}^{N} S_{osc} + \sum_{i=1}^{M} S_{OR}\right) \qquad (1)$$

where $S_{R_o}$ and $S_{R_F}$ is the entropy produced by the resistors R$_o$ and R$_F$, respectively. $S_{OSC}$ is the entropy produced by the oscillators and can be assumed constant since the nominal entropy produced by an oscillator over a cycle is almost constant; thus, it is unlikely to affect the system evolution and computational dynamics. $S_{OR}$ is the entropy produced by the OR gates and depends on the output currents of the OR gates; if the output of an OR gate is 0, no current flows through the pull-down network as a unidirectional diode connected MOSFET is used at the output. On the other hand, if the output of an OR gate is logic 1, a current will flow from the pull-up network to the feedback circuit. Hence, maximization of entropy produced by the OR gates is favored when their outputs are at logic 1 i.e., $k_i = 1$. Now, considering $S_{OR}$ favors the maximization of $\sum_{i=1}^{M} k_i$, we will consider the other two remaining terms in equation (1) to evaluate whether they favor such maximization. Equation (1) can subsequently be reduced to:

$$\frac{\partial S}{\partial t} = \frac{\partial}{\partial t} \left( \sum_{i=1}^{M} k_i S_{R_o} + \sum_{j=1}^{N} S_{R_F} \right) \tag{2}$$

The first term in equation (2) describing the entropy produced by the output resistors (R$_o$) can be expressed as:

$$\frac{\partial}{\partial t} \left( \sum_{i=1}^{M} k_i S_{R_o} \right) = \sum_{i=1}^{M} \left[ k_i \frac{\partial}{\partial t} \left( S_{R_o} \right) + S_{R_o} \frac{\partial}{\partial t} (k_i) \right] \tag{3}$$

The entropy produced by a resistor R passing a current $i$ over time $t$ and given off to the environment which is maintained at an ambient temperature $T_a$ is given by: $\frac{\Delta Q}{T_a} = \frac{\int_0^t i^2 R dt}{T_a}$ [53, 54] ($\Delta Q$ is the energy (dissipated as heat here) produced by the system). Therefore, the average entropy produced by the output resistor R$_o$ (over the nominal time period T of the oscillator) can be expressed as: $S_{R_o} = \frac{\int_0^T I_{o,av}^2 R_o \, dt}{T_a}$ ($I_{o,av}$ is the average current over the time period T) and the corresponding entropy generation rate is $\frac{I_{o,av}^2 R_o}{T_a}$. Furthermore, the second term $S_{R_o} \frac{\partial}{\partial t}(k_i)$ over the same period is close to zero since the OR gates undergo an even number (0 or 2) of transitions during this time period (one low-to-high and one high-to-low transition, if two transitions occur) which cancel each other (assuming that the positive and the negative transitions are symmetric); details in Supplement S1A. Thus, the first term in equation (2) evolves to,

$$\frac{\partial}{\partial t}\left(\sum_{i=1}^M k_i S_{R_o}\right) \cong \sum_{i=1}^M k_i \frac{I_{o,av}^2 R_o}{T_a} \equiv \frac{I_{o,av}^2 R_o}{T_a} \sum_{i=1}^M k_i \qquad (4)$$

To analyze the second term in equation (2) describing the entropy produced by the feedback resistors R_F, we apply Kirchhoff's current law (KCL) at the output node (Y). Assuming that the oscillator amplitude $v_j$ is much smaller than the voltage $V_Y$ at the output node, the average current through the feedback resistors $I_{R_F}$ can be considered almost equal, and can be expressed as $I_{R_F,av} \cong \frac{(\sum_{i=1}^M k_i) I_{o,av}}{N}$. Consequently, it can be shown that the entropy generation rate in the feedback resistors is $\frac{\partial}{\partial t}\left(\sum_{j=1}^N S_{R_F}\right) \cong \frac{R_F}{T_a}\sum_{j=1}^N \left(\frac{I_{o,av}}{N}\right)^2 \left(\sum_{i=1}^M k_i\right)^2$ (details shown in Supplement S1B). Substituting these terms, equation (2) evolves to:

$$\frac{\partial S}{\partial t} \cong \frac{(I_{o,av}^2 R_o)}{T_a}\sum_{i=1}^M k_i + \frac{R_F}{T_a}\sum_{j=1}^N \left(\frac{I_{o,av}}{N}\right)^2 \left(\sum_{i=1}^M k_i\right)^2 \qquad (5)$$

It can be observed from equation (5) that the system will aim to maximize the entropy production rate by maximizing $\sum_{i=1}^M k_i$, or, in other words, the number of satisfied clauses.

**System dynamics and analytical model.** To describe the system dynamics, we first calculate the output voltage $V_Y$ (at node Y) using Kirchhoff's current law:

$$\sum_{i=1}^{M} k_i I_{o,av} = \sum_{j=1}^{N} \frac{V_Y - v_j}{R_F} \qquad (6)$$

$$V_Y = \frac{1}{N} \left( R_F I_{o,av} \sum_{i=1}^{M} k_i + \sum_{j=1}^{N} v_j \right) \qquad (7)$$

Since the amplitude of the oscillators is small, and N<<M (usually the case for hard MaxSAT problems), the output node voltage can be simplified and expressed as:

$$V_Y \approx \frac{R_F}{N} I_{o,av} \sum_{i=1}^{M} k_i \cong K \sum_{i=1}^{M} k_i \qquad (8)$$

where $K$ can be considered as a constant representing the feedback. Furthermore, $V_Y$ can be considered as an external perturbation signal for the oscillators. The resulting dynamics of the oscillators under influence of this external signal $V_Y(t)$ can be described by[55,56]:

$$\frac{d\alpha_j}{dt} = \vartheta \left( \omega t + \alpha_j(t) \right) \cdot V_Y(t) \quad (j = 1, 2, \dots N) \qquad (9)$$

where $\alpha_j$ is the phase deviation of the oscillator, and $\vartheta$ is the perturbation projection vector (PPV) of the oscillator, which details the output of the oscillator in response to an external perturbation[57,58]. Considering sinusoidal oscillators in this analysis (used here for simplicity), the PPV is sinusoidal as well. The resulting dynamics of the oscillators can be expressed as:

$$\frac{d\alpha_j}{dt} = A \sin \left( \omega t + \alpha_j(t) \right) \cdot \left( K \sum_{i=1}^{M} k_i \right) \quad (j = 1, 2, \dots N) \qquad (10)$$

Where, A is the amplitude of the PPV. Equation (10) can be solved using a standard

SDE (stochastic differential equation) with additive white noise[58].

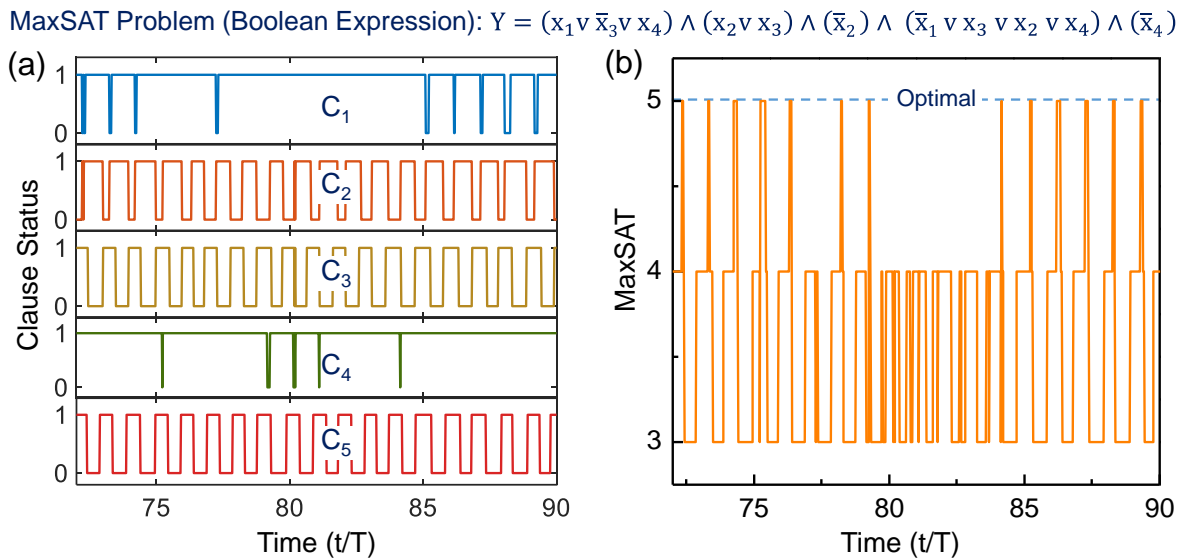

**Fig. 2| Solving MaxSAT using maximum entropy production principle in oscillator networks.** Evolution of **a,** status of the clauses; and **b,** total number of satisfied clauses, with time, for an illustrative Boolean expression in CNF format (N= 4; M=5) (shown in the figure). The oscillator dynamics are simulated using a standard stochastic differential equation (SDE) solver. Time is normalized to the time period (T) of the oscillator.

Fig. 2 shows a representative MaxSAT instance solved using the proposed model. Figs. 2a,b show the temporal evolution of the OR gate outputs ($\equiv$status of the clauses), and the total number of clauses satisfied, respectively. It can be observed that the computed solution agrees with the known optimal solution for the problem. In the following section, we test the proposed model on larger MaxSAT instances.

**Large MaxSAT Instances.** We consider various datasets from the 11[th] MaxSAT evaluation (Fig. 3). Figures 3a,b show the accuracy of the calculated solution and the projected time-to-compute for the model to converge to the optimal/near-optimal solution for 32 instances of the High-Girth (HG) 4-SAT dataset. The time-to-compute is measured in terms of T i.e., the number of oscillation periods, and reflects the time required for the

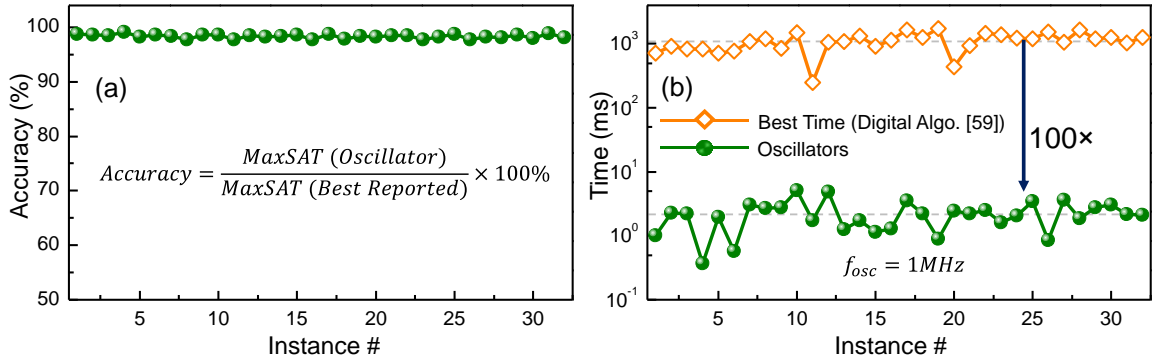

**Fig. 3| Evaluation of large MaxSAT instances. a,** Accuracy; and **b,** Projected computation time for (32) instances taken from the HG-4SAT benchmark. An oscillator frequency of 1MHz is assumed (see main text for details). **c,** Table showing accuracy (mean and best) and projected computation time (considering two different oscillator frequencies) for various MaxSAT datasets. All datasets are taken from 11th MaxSAT evaluation problems. A comparison of the computation time with digital algorithms is also shown. The best reported time for solving a particular dataset from a set of fourteen digital algorithms is considered [59].

| Dataset | No. of Instance | N | M | Average Accuracy | Maximum Accuracy | Average Computation Time (Oscillators) | | | Average Computation Time (Digital Algo.*) [59] |
|---|---|---|---|---|---|---|---|---|---|
| | | | | | | T | @ 1MHz | @ 50kHz | |
| High-Girth-4SAT | 7 | 100 | 900 | 98.65% | 99.1% | 1600 | 1.6 ms | 32 ms | 804.2 ms |
| High-Girth-4SAT | 25 | 150 | 1350 | 98.27% | 98.81% | 2430 | 2.43 ms | 48.6 ms | 1195.6 ms |
| High-Girth-3SAT | 25 | 250 | 1000 | 96.1% | 96.88% | 2200 | 2.2 ms | 44 ms | 1250.4 ms |
| High-Girth-3SAT | 25 | 300 | 1200 | 95.89% | 96.48% | 2550 | 2.55 ms | 51 ms | 1538 ms |
| Abrame-Habet-3SAT | 10 | 110 | 700 | 97.8% | 98.7% | 1280 | 1.28 ms | 25.6 ms | 874 ms |
| Abrame-Habet-3SAT | 10 | 110 | 800 | 97.7% | 98.5% | 2220 | 2.22 ms | 44.4 ms | 1086 ms |

*smallest reported time for the best-known solution.

model to converge to the solution. Subsequently, we project the actual time required by considering two different oscillator frequencies (50kHz and 1MHz). We do note that the physical platform that implements this model will have additional sources of delay. However, considering the relatively low oscillator frequencies, it is expected that the oscillator time period will be the primary bottleneck i.e., all other delays will be significantly lesser than T. Furthermore, this is corroborated in the following sections where the simulated hardware platform (with oscillator frequency ~1MHz) converges to the solution in ~8µs. The table in Fig. 3c shows the results for instances from various databases, where it can be observed that the oscillators facilitate high-quality near-optimal solutions with an minimum (average) accuracy of ~96%.

Unlike smaller instances, we observe that for instances with large number of clauses, while the system converges to a near-optimal solution very fast, convergence to an optimal solution is not guaranteed (or the system may require an inordinately large time), due to getting trapped in local maxima. In fact, after convergence to a near-optimal solution, the system exhibits an exponential increase in computation time to improve the solution as illustrated in the supplement S2. This is not surprising since this classical dynamical system implementation does not reduce the fundamental complexity of the problem. Consequently, to improve the quality of solutions without incurring a severe time penalty, we develop a simple heuristic post-processing technique based on an iterative approach. The proposed appraoch along with the corresponding improvement in the solutions and the time penalty for the post-processing has been described in supplement (S3). It is noted that Fig. 3 shows the results after post-processing (the results before post-processing are shown Supplement S3).

**Experimental Validation**. We now focus on the experimental validation of the proposed model on small MaxSAT instances, as well as develop a compute-in-memory (CiM)-based implementation in the following section. We first evaluate, experimentally, the same SAT expression considered in Fig. 2 (also shown in Fig. 4a in the CNF format). To realize the experimental circuit, we use LC oscillators (details of the design are described in Supplement S4) and implement the logical OR gates using an IC CD4072B. Further, $R_o$ (=10 KΩ, here) and $R_F$ (=10 KΩ, here) are adjusted to achieve the desired level of feedback; a very strong feedback locks the oscillators in-phase to the output and inhibits the dynamics desired for computation. Figs. 4b,c show the time domain evolution of the (buffered) oscillator outputs and the clauses (i.e., OR gate outputs), respectively. The evolution of the total number of clauses satisfied, shown in Fig. 4d, corroborates the ability of the oscillators to correctly solve the problem. Furthermore, we experimentally evaluate six example Boolean expressions $P_1$-$P_6$ ($P_1$-$P_3$: unsatisfiable; $P_4$-$P_6$ satisfiable; the Boolean expressions are detailed in Supplement S5) as shown in Fig. 4e. Optimal solutions are observed in all the cases. Moreover, we also note that for the unsatisfiable problems, the system never exhibits an incorrect solution i.e., shows an unsatisfiable problem as being satisfiable.

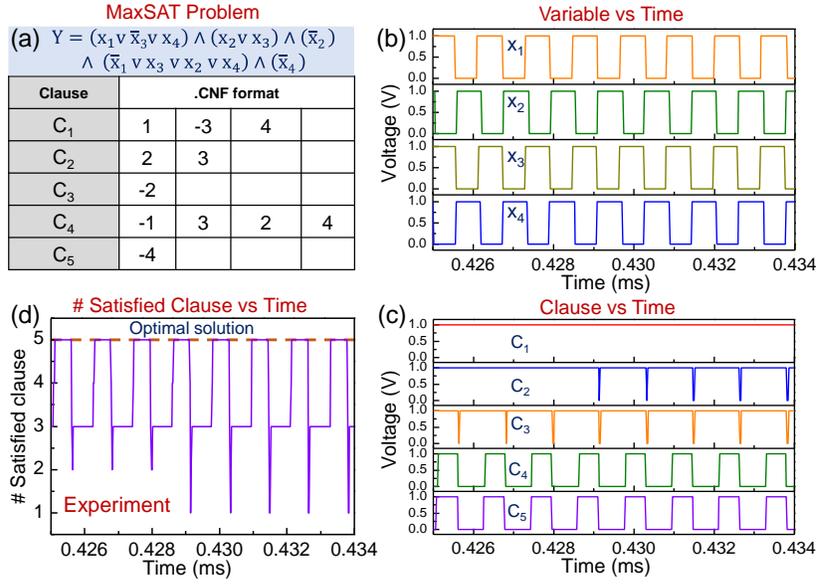

**Fig. 4| Experimental validation of the proposed model. a,** Representative MaxSAT problem expressed in the conjunctive normal format (same as the instance considered in Fig. 2). Experimentally measured temporal evolution of the: **b,** Buffered output of the oscillators (representing variables); **c,** Clauses, represented by the output of the OR gates; **d,** Total number of satisfied clauses. **e,** Experimentally measured MaxSAT solutions for six illustrative problems using the above oscillator approach; $P_1$-$P_3$ problems are unsatisfiable; $P_4$-$P_6$ are satisfiable. The temporal evolution of the total number of clauses is shown. The MaxSAT solutions (experimentally measured and optimal) are shown as the ratio of the number of satisfied clauses to the total number of clauses. Optimal solutions are observed in all cases.

**Compute-in-Memory (CiM) implementation.** While the circuit and experiments described above elucidate the conceptual operation and implementation of the oscillator-based MaxSAT solver, here, we propose a novel non-volatile memory (NVM)-based CiM approach to program and dynamically evaluate the clauses directly in memory (Fig. 5a). The CiM module is realized as a NOR array of 1T1R cells, where the NVM element is implemented using a Cu/HfO$_2$-based conducting bridge RAM (CBRAM) (Supplement S6, Fig. S6d); the fabrication process and detailed electrical characteristics of these structures have been experimentally demonstrated in prior work[60]. The choice of this material system is motivated by the high $R_{OFF}/R_{ON}$ ratio (>$10^5$) provided by CBRAM cell though the challenges of the material have been addressed in the following paragraph.

Each column of the array (i.e., bit line) represents a clause whereas each row (i.e., horizontal word line (WL)) represents a variable (a separate WL is used for the normal ($x_i$) and the negated form, $\bar{x}_i$) and is connected to the output of an oscillator (through buffers). A tri-state inverter is used to generate the variable's negated form with E (Enable) =1(0) during the execution (programming) phase. The 1T1R cell performs the dual function of representing if a variable (or its negated form) appears (or is absent) in a clause (using the state of the CBRAM) as well as assigning a truth value to the variables during execution of the MaxSAT problem (using the state of the FET). The operation of the hardware platform can be considered in two stages: (a) *Programming the clauses*:

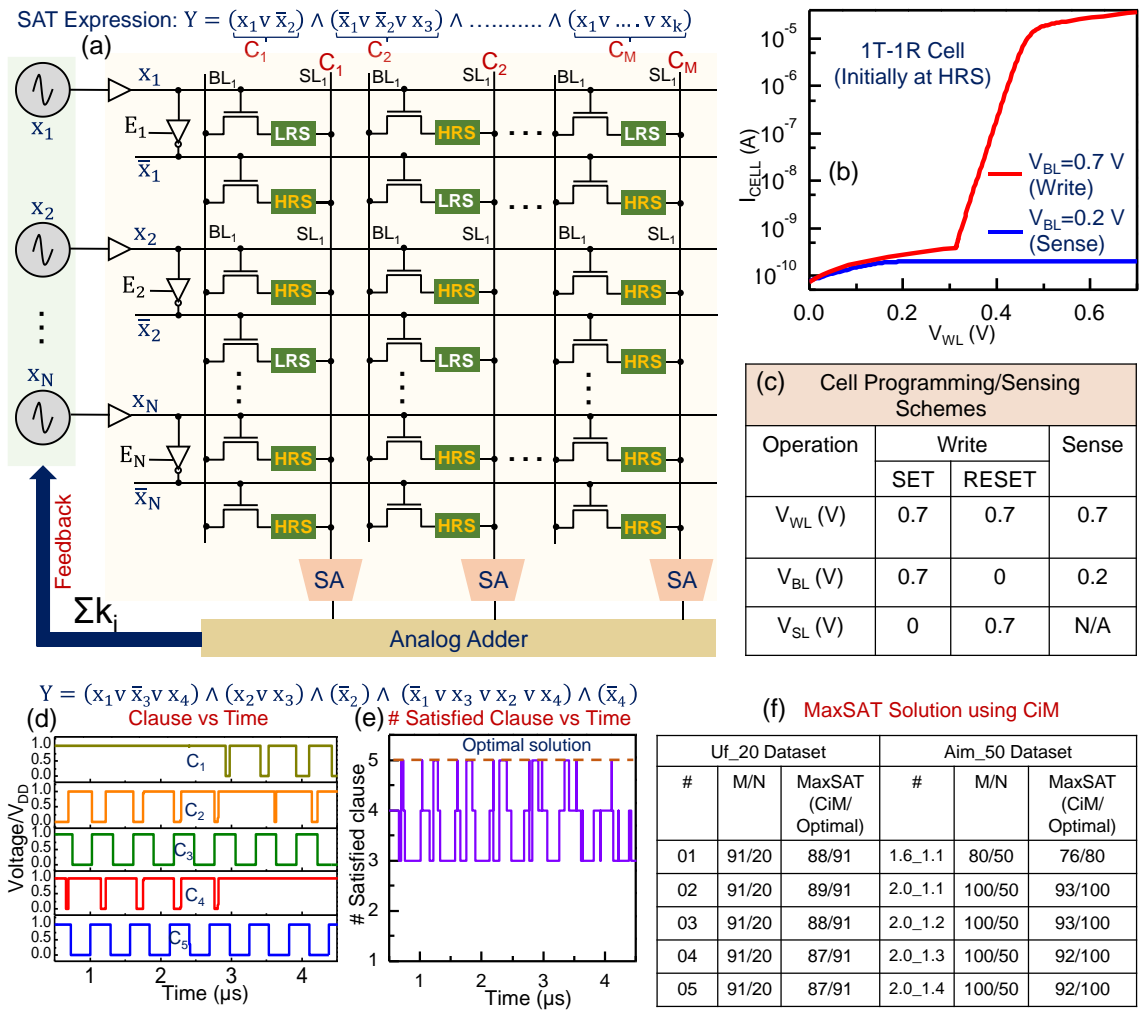

**Fig. 5| Compute-in-Memory (CiM)-based hardware implementation**. **a,** Schematic of the CiM-based hardware architecture to implement the MaxSAT solver. A MaxSAT problem of N variable & M clauses will require N oscillators, and (2N) x M NVM (non-volatile memory) array in the CiM block (2xN: variable & its negated form). **b,** Transfer characteristics of the 1T-1R cell (initially in HRS) used in the CiM block. **c,** Voltage levels used during programming of the memory cells (to represent the clauses) as well as during MaxSAT execution (sense). Simulated results for a representative MaxSAT problem (N=4, M=5). Evolution of **d,** clauses (output of sense amplifier); and **e,** MaxSAT solution (total number of sense amplifiers with high output), as a function of time. **f,** MaxSAT solutions computed using the CiM hardware for various problem instances from the Uf and the Aim datasets.

The NV CBRAM is initially programmed to represent if a variable is present (or absent) in a clause- when a variable appears in a clause, the CBRAM in the corresponding cell is set into the LRS (low resistance state); when a variable does not appear, the CBRAM is set into the HRS (high resistance state) which effectively "eliminates" the corresponding variable from the clause. For instance, it can be observed in Fig. 5a that since (only) $x_1$ and $\bar{x}_2$ appear in clause C$_1$ of the SAT expression, the corresponding CBRAMs are in the LRS whereas all other CBRAMs in the column are in the HRS. It is noted that the CBRAM needs to be programmed only once in every execution. Furthermore, although retention is a well-known challenge in the CBRAMs[61], the CBRAM in the CiM module, unlike in its application as a non-volatile memory, needs to retain the state only for the duration of the problem execution (~ms). (b) *MaxSAT execution*: During execution, the state of the FET which evolves with the oscillator output assigns the true value to the enabled cells (i.e., cells with CBRAM in LRS): when the FET is ON (V$_{GS}$=V$_{DD}$), it represents '1'; when OFF (V$_{GS}$=0), it represents '0'. It is noted that, BL bias is small enough to ensure that the states of the CBRAMS are not disturbed. The sense amplifier (SA) for each column ($\equiv$clause) reads and outputs a '1' if at least one enabled cell in the column is 1 (i.e., FET is ON); in other words, the SA effectively computes a logical 'OR' of the variables in the clause. Subsequently, the outputs of the SAs are summated using an analog adder and is used as feedback to all the oscillators; the feedback is proportional to the number of satisfied clauses ($\sum_1^M k_i$), as dictated by the computational model.

We simulate this circuit using the 14 nm predictive technology model (PTM)[62] for the transistor, and a phenomenological SPICE-based compact model[63] for the Cu/HfO$_2$-based CBRAM. Fig. 5b shows the resulting I$_{DS}$-V$_{GS}$ characteristics of the cell with the NVM element in the LRS (low resistance state) and HRS (high resistance state), along with the program/sense parameters shown in Fig. 5c. Additional details of the circuit implementation along with the programming and sensing schemes for the memory cells have been discussed in the supplement S6 and S7. Figure 5d,e show the temporal evolution of the oscillators and the clauses (i.e., the output of the SAs) for the representative problem considered in Fig. 2; the system is simulated for a total period of 8 µs. Furthermore, we also evaluate the performance of the proposed implementation for 10 instances of the Uf and Aim database[64] where (Fig. 5f) it can be observed that the oscillators compute high-quality solutions to the MaxSAT problem with an average accuracy of 96% (without requiring post processing); optimal solutions are obtained after postprocessing in all the instances.

**Conclusion**

In summary, this work develops the foundation for using the physics of oscillators, interacting through an engineered feedback, to solve to the non-quadratic MaxSAT problem. We show, and corroborate with prototype experiments, that the combinatorial function of the MaxSAT problem can be directly mapped to the maximization of the entropy generation rate of the proposed oscillator system. Furthermore, we also demonstrate a compute-in-memory architecture for the hardware implementation of the proposed model. Our work marks an important step towards enabling application-specific non-von Neumann hardware that exploits new computing models to solve hard problems. While the present work demonstrates the computing model, future work on system optimization aspects such as design and incorporation of an annealing schedule, optimization of the feedback, role of oscillator realization can further help optimize the computational performance.

**References**

1. Calude, C.S. Unconventional computing: A brief subjective history. In *Advances in Unconventional Computing*, Springer, Cham, 855-864 (2017).

2. Chen, A., Datta, S., Hu, X.S., Niemier, M.T., Rosing, T.S. & Yang, J.J. A survey on architecture advances enabled by emerging beyond-CMOS technologies. *IEEE Design & Test* **36**, 46-68 (2019).

3. Johnson, M. W. *et al.* Quantum annealing with manufactured spins. *Nature* **473**, 194-198 (2011).

4. Pervaiz, A. Z., Ghantasala, L. A., Camsari, K. Y. & Datta, S. Hardware emulation of stochastic p-bits for invertible logic. *Scientific Reports* **7**, 1-13 (2017).

5. Borders, W. A., Pervaiz, A. Z., Fukami, S., Camsari, K. Y., Ohno, H. & Datta, S. Integer factorization using stochastic magnetic tunnel junctions. *Nature* **573**, 390-393 (2019).

6. Mostafa, H., Müller, L. K. & Indiveri, G. An event-based architecture for solving constraint satisfaction problems. *Nature Communications* **6**, 1-10 (2015).

7. Nguyen, A. H. N., Aono, M. & Hara-Azumi, Y. FPGA-Based Hardware/Software Co-Design of a Bio-Inspired SAT Solver. *IEEE Access* **8**, 49053-49065 (2020).

8. Maximum satisfiability problem retrieved on Feb. 3, 2021. At https://en.wikipedia.org/wiki/Maximum_satisfiability_problem

9. Si, X., Zhang, X., Grigore, R. & Naik, M. Maximum satisfiability in software analysis: Applications and techniques. In *International Conference on Computer Aided Verification*, Springer, Cham, 68-94, (2017).

10. Berg, O.J., Hyttinen, A.J. & Jarvisalo, M.J. Applications of MaxSAT in data analysis. In *Proceedings of Pragmatics of SAT 2015 and 2018* (2019).

11. Walter, R., Zengler, C. & Küchlin, W. Applications of MaxSAT in Automotive Configuration. In *Configuration Workshop* **1**, 21 (2013).

12. Safarpour, S., Mangassarian, H., Veneris, A., Liffiton, M.H. & Sakallah, K.A. Improved design debugging using maximum satisfiability. In *Formal Methods in Computer Aided Design (FMCAD'07)*, 13-19, doi: 10.1109/FAMCAD.2007.26 (IEEE, 2007).

13. Demirovic, E., Musliu, N. & Winter, F. Modeling and solving staff scheduling with partial weighted maxSAT. *Annals of operations research* **275**, 79-99 (2019).

14. Xu, H., Rutenbar, R.A. & Sakallah, K. sub-SAT: a formulation for relaxed boolean satisfiability with applications in routing. *IEEE Transactions on Computer-Aided Design of Integrated Circuits and Systems* **22**, 814-820 (2003).

15. Rintanen, J., Heljanko, K. & Niemelä, I. Planning as satisfiability: parallel plans and algorithms for plan search. *Artificial Intelligence* **170**, 1031-1080 (2006).

16. Cooper, G.F. The computational complexity of probabilistic inference using Bayesian belief networks. *Artificial intelligence* **42**, 393-405 (1990).

17. Hopfield, J.J. Neural networks and physical systems with emergent collective computational abilities. *Proceedings of the National Academy of Sciences* **79**, 2554-2558 (1982).

18. Siegelmann, H.T. & Fishman, S. Analog computation with dynamical systems. *Physica D: Nonlinear Phenomena* **120**, 214-235 (1998).

19. Crutchfield, J.P., Ellison, C.J., James, R.G. & Mahoney, J.R. Synchronization and control in intrinsic and designed computation: An information-theoretic analysis of competing models of stochastic computation. *Chaos: An Interdisciplinary Journal of Nonlinear Science* **20**, 037105 (2010).

20. Legenstein, R. & Maass, W. What makes a dynamical system computationally powerful. *New directions in statistical signal processing: From systems to brain*, 127-154 (2007).

21. Stepney, S. Nonclassical Computation-A Dynamical Systems Perspective. Handbook of natural computing **2** (2012).

22. MacLennan, B.J. The promise of analog computation. *International Journal of General Systems* **43**, 682-696 (2014).

23. Hasler, J. Opportunities in physical computing driven by analog realization. In *2016 IEEE International Conference on Rebooting Computing (ICRC)*, 1-8, doi: 10.1109/ICRC.2016.7738680 (IEEE, 2016).

24. Fukami, S., Borders, W.A., Pervaiz, A.Z., Camsari, K.Y., Datta, S. & Ohno, H. Probabilistic computing based on spintronics technology. In *2020 IEEE Silicon Nanoelectronics Workshop (SNW)*, 21-22, doi: 10.1109/SNW50361.2020.9131622 (IEEE, 2020).

25. Marandi, A., Wang, Z., Takata, K., Byer, R.L. & Yamamoto, Y. Network of time-multiplexed optical parametric oscillators as a coherent Ising machine. *Nature Photonics* **8**, 937-942 (2014).

26. Di Ventra, M. & Traversa, F.L. Perspective: Memcomputing: Leveraging memory and physics to compute efficiently. *Journal of Applied Physics* **123**, 180901 (2018).

27. Hoppensteadt, F. C. & Izhikevich, E. M. Synaptic organizations and dynamical properties of weakly connected neural oscillators. *Biological cybernetics* **75**, 117-127 (1996).

28. Csaba, G. & Porod, W. Coupled oscillators for computing: A review and perspective. *Applied Physics Reviews* **7**, 011302 (2020).

29. Csaba, G., Raychowdhury, A., Datta, S. & Porod, W. Computing with coupled oscillators: Theory, devices, and applications. In *2018 IEEE International Symposium on Circuits and Systems (ISCAS)*, 1-5, doi: 10.1109/ISCAS.2018.8351664 (IEEE, 2018).

30. Nikonov, D. E., Csaba, G., Porod, W., Shibata, T., Voils, D., Hammerstrom, D. *et al.* Coupled-oscillator associative memory array operation for pattern recognition. *IEEE Journal on Exploratory Solid-State Computational Devices and Circuits* **1**, 85-93 (2015).

31. Fang, Y., Yashin, V. V., Chiarulli, D. M. & Levitan, S. P. A simplified phase model for oscillator based computing. In *2015 IEEE Computer Society Annual Symposium on VLSI*, 231-236, doi: 10.1109/ISVLSI.2015.44 (IEEE, 2015).

32. Vodenicarevic, D., Locatelli, N., Araujo, F. A., Grollier, J. & Querlioz, D. A nanotechnology-ready computing scheme based on a weakly coupled oscillator network. *Scientific Reports* **7**, 1-13 (2017).

33. Nikonov, D. E., Kurahashi, P., Ayers, J. S., Lee, H. J., Fan, Y. & Young, I. A. A Coupled CMOS Oscillator Array for 8ns and 55pJ Inference in Convolutional Neural Networks. Preprint at https://arxiv.org/abs/1910.11803 (2019).

34. Yogendra, K., Liyanagedera, C., Fan, D., Shim, Y. & Roy, K. Coupled spin-torque nano-oscillator-based computation: A simulation study. *ACM Journal on Emerging Technologies in Computing Systems (JETC)* **13**, 1-24 (2017).

35. Malagarriga, D., García-Vellisca, M. A., Villa, A. E., Buldú, J. M., García-Ojalvo, J. & Pons, A. J. Synchronization-based computation through networks of coupled oscillators. *Frontiers in computational neuroscience* **9**, 97 (2015).

36. Romera, M. *et al.* Vowel recognition with four coupled spin-torque nano-oscillators. *Nature* **563**, 230-234 (2018).

37. Molnar, B., Molnar, F., Varga, M., Toroczkai, Z. & Ercsey-Ravasz, M. A continuous-time MaxSAT solver with high analog performance. *Nature Communications* **9**, 1-12 (2018).

38. Traversa, F.L., Cicotti, P., Sheldon, F. & Di Ventra, M. Evidence of exponential speed-up in the solution of hard optimization problems. *Complexity* (2018).

39. Mallick, A., Bashar, M. K., Truesdell, D. S., Calhoun, B. H., Joshi, S. & Shukla, N. Using synchronized oscillators to compute the maximum independent set. *Nature Communications* **11**, 1-7 (2020).

40. Wang, T. & Roychowdhury, J. OIM: Oscillator-based Ising machines for solving combinatorial optimisation problems. In *International Conference on Unconventional Computation and Natural Computation*, 232-256, Springer, Cham (2019).

41. Chou, J., Bramhavar, S., Ghosh, S. & Herzog, W. Analog coupled oscillator based weighted Ising machine. *Scientific Reports* **9**, 1-10 (2019).

42. Dutta, S., Khanna, A., Gomez, J., Ni, K., Toroczkai, Z. & Datta, S. Experimental Demonstration of Phase Transition Nano-Oscillator Based Ising Machine. In *2019 IEEE International Electron Devices Meeting (IEDM)*, 37-8, doi: 10.1109/IEDM19573.2019.8993460, (IEEE, 2019).

43. Ahmed, I., Chiu, P. W. & Kim, C. H. A Probabilistic Self-Annealing Compute Fabric Based on 560 Hexagonally Coupled Ring Oscillators for Solving Combinatorial Optimization Problems. In *2020 IEEE Symposium on VLSI Circuits*, 1-2, doi: 10.1109/VLSICircuits18222.2020.9162869, (IEEE, 2020).

44. Bashar, M.K., Mallick, A., Truesdell, D.S., Calhoun, B.H., Joshi, S. & Shukla, N. Experimental Demonstration of a Reconfigurable Coupled Oscillator Platform to Solve the Max-Cut Problem. *IEEE Journal on Exploratory Solid-State Computational Devices and Circuits* **6**, 116-121, doi: 10.1109/JXCDC.2020.3025994, (2020).

45. Dutta, S., Khanna, A. & Datta, S. Understanding the Continuous-Time Dynamics of Phase-Transition Nano-Oscillator-based Ising Hamiltonian Solver. *IEEE Journal on Exploratory Solid-State Computational Devices and Circuits* **6**, 155-163 (2020).

46. Parihar, A., Shukla, N., Jerry, M., Datta, S. & Raychowdhury, A. Vertex coloring of graphs via phase dynamics of coupled oscillatory networks. *Scientific Reports* **7**, 1-11 (2017).

47. Mallick, A., Bashar, M. K., Truesdell, D. S., Calhoun, B. H., Joshi, S. & Shukla, N. Graph Coloring using Coupled Oscillator-based Dynamical Systems. In *2021 IEEE International Symposium on Circuits and Systems (ISCAS)*, 1-5, doi: 10.1109/ISCAS51556.2021.9401188 (IEEE, 2021).

48. Vaidya, J., Bashar, M.K. & Shukla, N. Using Noise to Augment Synchronization among Oscillators. *Scientific Reports* **11**, 4462 (2021).

49. Gabor, T., Zielinski, S., Feld, S., Roch, C., Seidel, C., Neukart, F., Galter, I., Mauerer, W. & Linnhoff-Popien, C. Assessing solution quality of 3SAT on a quantum annealing platform. In *International Workshop on Quantum Technology and Optimization Problems*, 23-35, Springer, Cham, doi: 10.1007/978-3-030-14082-3_3 (2019).

50. Reformulating a Problem, retrieved on August 9, 2021. Available at: https://docs.dwavesys.com/docs/latest/_downloads/c165bdc8f07ff4c9d142f3e07881814f/09-1171B-E_Developer_Guide_Problem_Solving_Handbook.pdf

51. Jackson, P. & Sheridan, D. Clause form conversions for boolean circuits. In *International Conference on Theory and Applications of Satisfiability Testing*, 183-198, Springer, Berlin, Heidelberg (2004).

52. Martyushev, L.M. & Seleznev, V.D. Maximum entropy production principle in physics, chemistry and biology. *Physics Reports* **426**, 1-45 (2006).

53. Christen, T. Application of the maximum entropy production principle to electrical systems. *Journal of Physics D: Applied Physics* **39**, 4497 (2006).

54. Zupanovic, P., Juretic, D. & Botric, S. Kirchhoff's loop law and the maximum entropy production principle. *Physical Review E* **70**, 056108 (2004).

55. Demir, A., Mehrotra, A. & Roychowdhury, J. Phase noise in oscillators: A unifying theory and numerical methods for characterization. *IEEE Transactions on Circuits and Systems I: Fundamental Theory and Applications* **47**, 655-674 (2000).

56. Srivastava, S. & Roychowdhury, J. Analytical equations for nonlinear phase errors and jitter in ring oscillators. *IEEE Transactions on Circuits and Systems I: Regular Papers* **54**, 2321-2329 (2007).

57. Bhansali, P. & Roychowdhury, J. Gen-Adler: The generalized Adler's equation for injection locking analysis in oscillators. In *2009 Asia and South Pacific Design Automation Conference*, 522-527, doi: 10.1109/ASPDAC.2009.4796533 (IEEE, 2009).

58. Wang, T. & Roychowdhury, J. Oscillator-based ising machine. Preprint at https://arxiv.org/abs/1709.08102 (2017).

59. Max-SAT Benchmark, retrieved on Feb. 9, 2021. At http://maxsat.ia.udl.cat/detailed/incomplete-ms-random-table.html

60. Shukla, N., Ghosh, R.K., Grisafe, B. & Datta, S. Fundamental mechanism behind volatile and non-volatile switching in metallic conducting bridge RAM. In *2017 IEEE International Electron Devices Meeting (IEDM)*, 4.3.1-4.3.4, doi: 10.1109/IEDM.2017.8268325 (IEEE, 2017).

61. Jameson, J.R., Blanchard, P., Cheng, C., Dinh, J., Gallo, A., Gopalakrishnan, V., Gopalan, C., Guichet, B., Hsu, S., Kamalanathan, D. & Kim, D. Conductive-bridge memory (CBRAM) with excellent high-temperature retention. In *2013 IEEE International Electron Devices Meeting*, 30-1, doi: 10.1109/IEDM.2013.6724721 (IEEE, 2013).

62. Latest Models, retrieved on March 5, 2021. At http://ptm.asu.edu

63. Aziz, A., Jao, N., Datta, S. & Gupta, S. K. Analysis of functional oxide based selectors for cross-point memories. *IEEE Transactions on Circuits and Systems I: Regular Papers* **63**, 2222-2235 (2016).

64. SATLIB - Benchmark Problems, retrieved on April 1, 2021. At https://www.cs.ubc.ca/~hoos/SATLIB/benchm.html

## Author contributions

M.K.B. and N.S. conceived the idea and developed the computational model. A.M. and J.V. performed the experiments. J.V., R.S.S.K. and M.K.B. performed the circuit simulation. M.K.B., S.A. and N.A. performed the memory computing simulation. M.K.B., C.L. and F.S. developed the post-processing approach. M.K.B. and N.S. wrote the manuscript. A.A., V.N., and N.S. supervised the study. All authors discussed the results and commented on the manuscript.

## Additional information

Correspondence and requests for materials should be addressed to N. S.

# Supplementary Information
# An Oscillator-based MaxSAT solver

Mohammad Khairul Bashar[1], Jaykumar Vaidya[1], Antik Mallick[1], R S Surya Kanthi[1], Shamiul Alam[2], Nazmul Amin[2], Chonghan Lee[3], Feng Shi[4], Ahmedullah Aziz[2], Vijaykrishnan Narayanan[3], Nikhil Shukla[1]*

[1]Department of Electrical and Computer Engineering, University of Virginia, Charlottesville, VA- 22904 USA

[2]Department of Electrical Engineering and Computer Science, University of Tennessee, Knoxville, TN 37996, USA

[3]Department of Computer Science and Engineering, Pennsylvania State University, University Park, State College, PA 16801, USA

[4]Department of Computer Science, University of California- Los Angeles, Los Angeles, CA 90024, USA

*Correspondence to ns6pf@virginia.edu

## S1. Entropy generation rate

**A. <u>No. of Transitions in the OR gate output</u>**. The OR gates (in Fig. 1) can undergo only an even number (0 or 2) of transitions in response to the oscillator inputs during a time period T. The output of the OR gate will be high if any of the oscillator inputs to the OR gate are high (i.e., $\geq V_{DD} / 2$); and the OR gate will output a zero if and only if all the inputs are low (i.e., $\leq V_{DD} / 2$). Since each oscillator is considered to have a nominal duty cycle of 50% i.e., the (buffered) output is symmetrically high and low for a time period of T/2 each, the OR gate output will be high for a minimum period of T/2. This also implies that the OR gate output can be low for a maximum time duration of T/2.

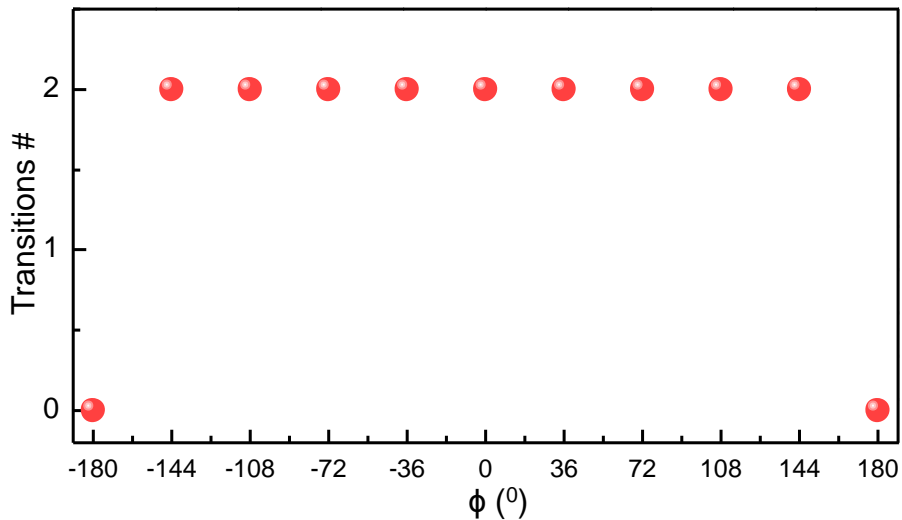

**Fig. S1| Transition characteristics of OR gates in response to oscillator inputs.** Number of transitions (high-to-low and low-to-high) per cycle (over time period T) in the output of a 2-input OR gate as a function of phase difference (ϕ) between the inputs. Square wave inputs with 50% duty cycle are considered here.

The OR output will be low for the maximum period of T/2 when all the oscillators are in-phase. Moreover, this also implies that within a given time period T, the transitions in the output of the OR gate will occur in pairs i.e., the high-to-low transition into the low logic state, followed by the low-to-high transition out of the low logic state. Furthermore, no transitions in the output state will occur if the oscillator phases are such that at least one oscillator is high during the time period T. For an OR gate with two oscillator inputs, this corresponds to the oscillators being out-of-phase; when the number of inputs is larger, a wide range of phase combinations can facilitate this condition. Figure S1 shows the number of possible transitions (in one cycle) in the output of a 2-input OR gate with respect to phase difference of the square wave inputs (having 50% duty cycle). It is observed that for a 2-input OR gate only an even number of transitions are possible. This is also applicable for OR gates with a larger number of inputs since the OR gate outputs zero for only one combination of inputs.

**B. Entropy generation rate in $R_F$.** The entropy generation rate in the feedback resistor $R_F$ over time $t$ can be expressed as:

$$\frac{\partial}{\partial t}\left(\sum_{j=1}^{N} S_{R_F}\right) = \frac{\partial}{\partial t}\sum_{j=1}^{N}\frac{\Delta Q}{T_a} = \frac{\partial}{\partial t}\sum_{j=1}^{N}\left(\frac{\int_0^t i_{R_F}^2 R_F dt}{T_a}\right) \tag{S1.1}$$

where,
$$i_{R_F} \cong \frac{1}{N}\sum_{i=1}^{M} k_i i_{R_o} \tag{S1.2}$$

Substituting S1.2 into S1.1,

$$\frac{\partial}{\partial t}\left(\sum_{j=1}^{N} S_{R_F}\right) = \frac{\partial}{\partial t}\sum_{j=1}^{N}\left(\frac{\int_0^t (\frac{1}{N}\sum_{i=1}^{M} k_i i_{R_o})^2 R_F dt}{T_a}\right) = \sum_{j=1}^{N}\frac{R_F}{N^2 T_a}\frac{\partial}{\partial t}\left(\int_0^t (\sum_{i=1}^{M} k_i)^2 i_{R_o}^2 dt\right) \tag{S1.3}$$

The integral in the right-hand side term in (S1.3) can be simplified as:

$$\frac{\partial}{\partial t}\left(\int (\sum_{i=1}^{M} k_i)^2 i_{R_o}^2 dt\right) = \frac{\partial}{\partial t}\left[(\sum_{i=1}^{M} k_i)^2 \int i_{R_o}^2 dt - \int \frac{\partial}{\partial t}(\sum_{i=1}^{M} k_i)^2 (\int i_{R_o}^2 dt)dt\right]$$

$$= 2(\sum_{i=1}^{M} k_i)\frac{\partial}{\partial t}(\sum_{i=1}^{M} k_i)(\int i_{R_o}^2 dt) + (\sum_{i=1}^{M} k_i)^2 \frac{\partial}{\partial t}(\int i_{R_o}^2 dt)$$

$$-\frac{\partial}{\partial t}\left[\int 2(\sum_{i=1}^{M} k_i)\frac{\partial}{\partial t}(\sum_{i=1}^{M} k_i)(\int i_{R_o}^2 dt)\,dt\right] \qquad (S1.4)$$

In one time period $T$, $k_i$ exhibits an even number of transitions (described in S1A) with the rise and fall transitions having almost the same rate. Hence, $\frac{\partial}{\partial t}(\sum_{i=1}^{M} k_i) = 0$, and subsequently, equation (S1.4) reduces to:

$$\frac{\partial}{\partial t}\left(\int (\sum_{i=1}^{M} k_i)^2 i_{R_o}^2 dt\right) = (\sum_{i=1}^{M} k_i)^2 \frac{\partial}{\partial t}\left(\int i_{R_o}^2 dt\right) \qquad (S1.5)$$

Further, substituting equation (S1.5) into equation (S1.3):

$$\frac{\partial}{\partial t}\left(\sum_{j=1}^{N} S_{R_F}\right) = \sum_{j=1}^{N}\frac{R_F}{N^2 T_a}(\sum_{i=1}^{M} k_i)^2 \frac{\partial}{\partial t}\left(\int_0^t i_{R_o}^2 dt\right) \qquad (S1.6)$$

To compute the entropy generation rate in $R_F$ over one time period (T), we consider that the average current flowing through $R_o$ over one time period is $I_{o,av}$. Thus, equation (S1.6) can be expressed as:

$$\frac{\partial}{\partial t}\left(\sum_{j=1}^{N} S_{R_F}\right) = \sum_{j=1}^{N}\frac{R_F}{N^2 T_a}(\sum_{i=1}^{M} k_i)^2 \frac{\partial}{\partial t}\left(\int_0^T I_{o,av}^2 dt\right) \qquad (S1.7)$$

$$= \sum_{j=1}^{N}\frac{R_F}{N^2 T_a}(\sum_{i=1}^{M} k_i)^2 \frac{I_{o,av}^2 T}{T}$$

$$= \frac{R_F}{T_a}\sum_{j=1}^{N}\left(\frac{I_{o,av}}{N}\right)^2 (\sum_{i=1}^{M} k_i)^2 \qquad (S1.8)$$

Equation (S1.8) describes the entropy generation rate in $R_F$ in one time period.

**S2. Temporal dependence of the quality of the MaxSAT solution**

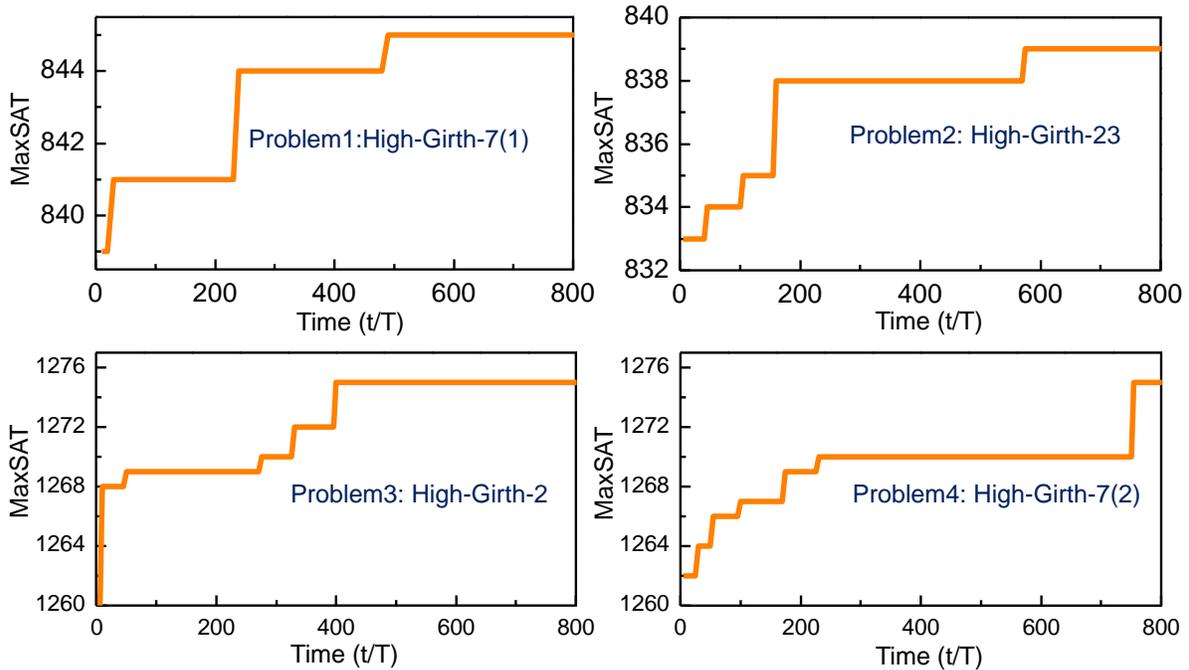

**Fig. S2| Accuracy vs. computation time trade-off.** Maximum number of clauses satisfied as a function of time for 4 representative problems from the HG-MaxSAT database (11th MaxSAT evaluation).

Figure S2 shows the maximum number of clauses satisfied as a function of time for four representative instances from the 11th MaxSAT evaluation database; the solution is measured after every 5 oscillation cycles. It can be observed that while the system converges to a near-optimal solution very fast, convergence (beyond that) towards an optimal solution is slow and shows an exponential-like trend. This is not unexpected since the problem is NP-hard.

## S3. Post-processing algorithm to improve MaxSAT solution in large instances

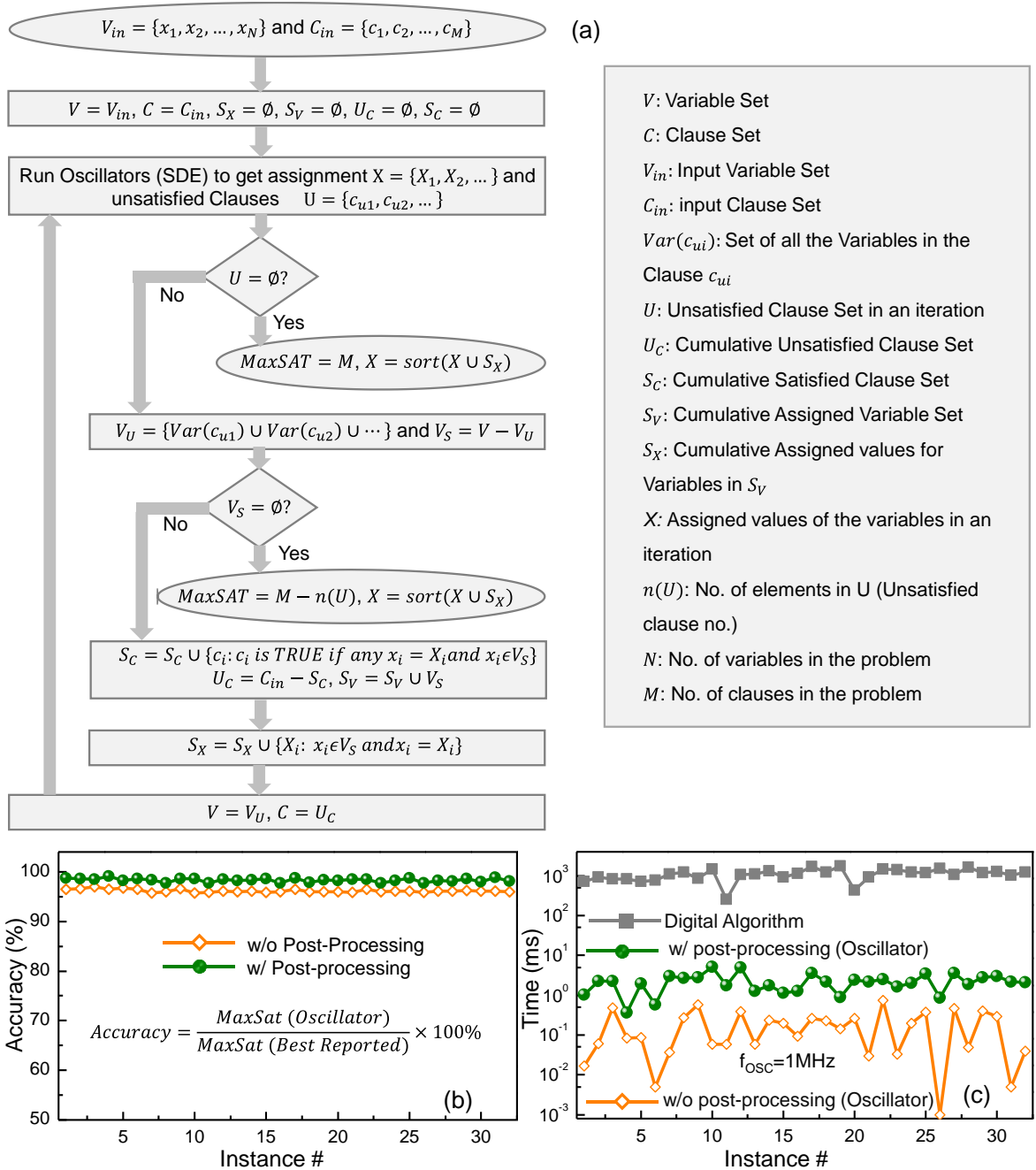

**Fig. S3| Post-processing scheme to improve MaxSAT solution. a,** Flowchart of the iterative post-processing algorithm used to improve the quality of the MaxSAT solution in large MaxSAT instances. **b,** Accuracy of-; and **c,** computation time for the MaxSAT solution before and after post-processing, evaluated for 32 problem instances from the HG-4-SAT dataset (11[th] Max-SAT evaluation).

Figure S3a shows the flow chart for the (heuristic) post-processing approach introduced here to improve the MaxSAT solution in large problem instances where the system may get stuck in a sub-optimal solution (corresponding to a local maximum in the high dimensional phase space). Using the solution obtained from the oscillators as the starting point, we identify the unsatisfied clauses and the (sub)set of variables ($U_V$; $n(U_V) = \eta$ ≤N) that appear in these clauses. Next, we identify all the satisfied clauses that would be satisfied irrespective of the truth assignments to $U_V$ (i.e., set of variables in the unsatisfied clauses). These clauses are eliminated from consideration and a reduced problem with $U_V$ variables and $U_C$ clauses is executed again using the oscillators. The iterative method (with reducing problem size in each run) is repeated until the number of variables cannot be reduced any further.

Figs. S3b,c illustrate the corresponding improvement in the solution as well as the time-penalty incurred with the post-processing scheme (evaluated on the 32 problem instances of the HG-4-SAT dataset (11[th] MaxSAT evaluation))[1,2].

## S4. Experimental setup for the oscillator based MaxSAT solver using OR gates

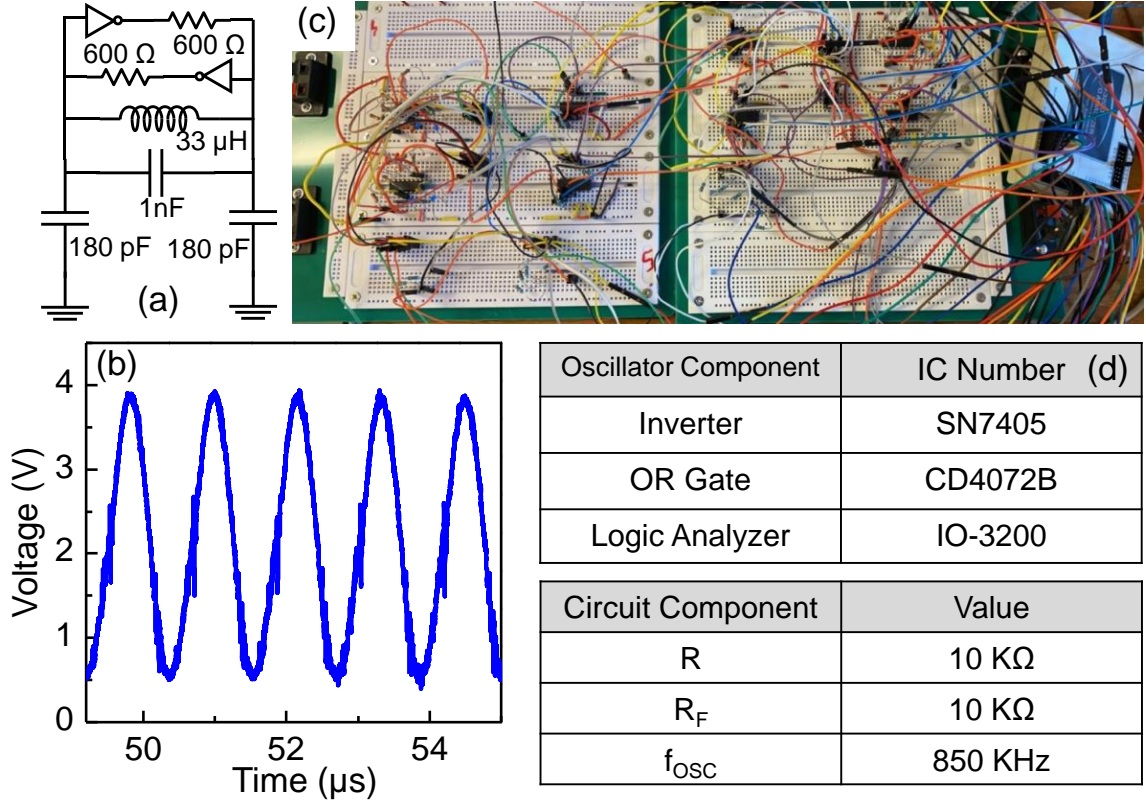

**Fig. S4| Experimental Setup. a,** Circuit schematic; and **b,** Time-domain waveform of the sinusoidal oscillator used in the experiment. **c,** Photograph of the experimental setup. **d,** Components and their values used to implement the oscillator-based MaxSAT solver.

The sinusoidal oscillator (Fig. S4a), representing a variable in the MaxSAT problem, is implemented using an inductor (33 µH), capacitor (1nF), trimmer capacitor (180 pF), cross-coupled inverters (from TI SN74HC05N ICs), and a 5V supply. Figure S4b shows the time domain output of the oscillator exhibiting a natural frequency of 850 KHz.

The clauses are implemented using OR-gates (CD4072B ICs). If the variable in the clause appears in the negated form, an inverting buffer (TI SN74HC05N ICs) is used. Fig. S4c shows a photograph of the experimental oscillator circuit implemented on a breadboard. The output of each OR gate is summated using a 10 kΩ resistor ($R_o$) while the feedback to the oscillators is also implemented using a 10 kΩ resistor ($R_F$) (parameters shown in Fig. S4d).

## S5. Boolean expressions for the experimentally measured MaxSAT instances

| Problem | Boolean Expression | MaxSAT (Experiment) | MaxSAT (Analytical) |
|---|---|---|---|
| $P_1$ | $Y = (x_1 \vee x_2) \wedge (x_1 \vee \bar{x}_2) \wedge (\bar{x}_1 \vee x_2) \wedge (\bar{x}_1 \vee \bar{x}_2)$ | 3/4 | 3/4 |
| $P_2$ | $Y = (x_1 \vee x_2 \vee x_4) \wedge (x_1 \vee \bar{x}_3) \wedge (x_1 \vee \bar{x}_4) \wedge$ $(\bar{x}_1 \vee x_3 \vee x_4) \wedge (\bar{x}_4) \wedge (\bar{x}_1 \vee x_4) \wedge (x_1)$ | 6/7 | 6/7 |
| $P_3$ | $Y = (x_2 \vee x_5 \vee x_6) \wedge (\bar{x}_4 \vee \bar{x}_5 \vee \bar{x}_6) \wedge (\bar{x}_1 \vee \bar{x}_3) \wedge$ $(x_2 \vee \bar{x}_6) \wedge (\bar{x}_2 \vee x_5) \wedge (x_1 \vee x_3) \wedge (x_3) \wedge (x_1 \vee \bar{x}_3)$ | 7/8 | 7/8 |
| $P_4$ | $Y = (x_1 \vee \bar{x}_3 \vee x_4) \wedge (\bar{x}_1 \vee x_2 \vee x_3 \vee x_4) \wedge (x_2 \vee x_3) \wedge (\bar{x}_2)$ $\wedge (\bar{x}_4) \wedge (x_1 \vee x_4) \wedge (x_3 \vee \bar{x}_4) \wedge (x_2 \vee \bar{x}_4)$ | 8/8 | 8/8 |
| $P_5$ | $Y = (x_1 \vee x_5 \vee \bar{x}_6) \wedge (x_2 \vee x_3) \wedge (\bar{x}_2 \vee \bar{x}_4 \vee x_5) \wedge$ $(\bar{x}_4 \vee x_6) \wedge (\bar{x}_1 \vee \bar{x}_5) \wedge (\bar{x}_3 \vee x_4 \vee x_5) \wedge$ $(x_2 \vee \bar{x}_3 \vee x_6) \wedge (x_2 \vee \bar{x}_5)$ | 8/8 | 8/8 |
| $P_6$ | $Y = (\bar{x}_1 \vee x_2 \vee x_4) \wedge (x_1 \vee x_3 \vee x_5) \wedge (x_2 \vee \bar{x}_3 \vee x_6) \wedge$ $(\bar{x}_2 \vee x_6) \wedge (x_1 \vee \bar{x}_6) \wedge (x_3 \vee \bar{x}_5 \vee x_6) \wedge$ $(\bar{x}_3 \vee \bar{x}_4) \wedge (\bar{x}_4 \vee \bar{x}_6) \wedge (\bar{x}_5 \vee \bar{x}_6)$ | 10/10 | 10/10 |

**Fig. S5| MaxSAT problem definitions.** Boolean expressions considered in Fig. 4 of the main text along with the MaxSAT solutions computed experimentally and analytically. The MaxSAT solutions (experimental and analytical) are shown as the ratio of the number of satisfied clauses to the total number of clauses. Optimal solutions are observed with both approaches.

Figure S5 shows the Boolean expressions for problems $P_1$-$P_6$ considered in Fig. 4 of the main text. Additionally, MaxSAT solutions computed experimentally and using the analytical approach have been shown in Fig. S5.

## S6. Hardware Implementation of oscillator-based MaxSAT solver with the compute-in-memory (CiM) module

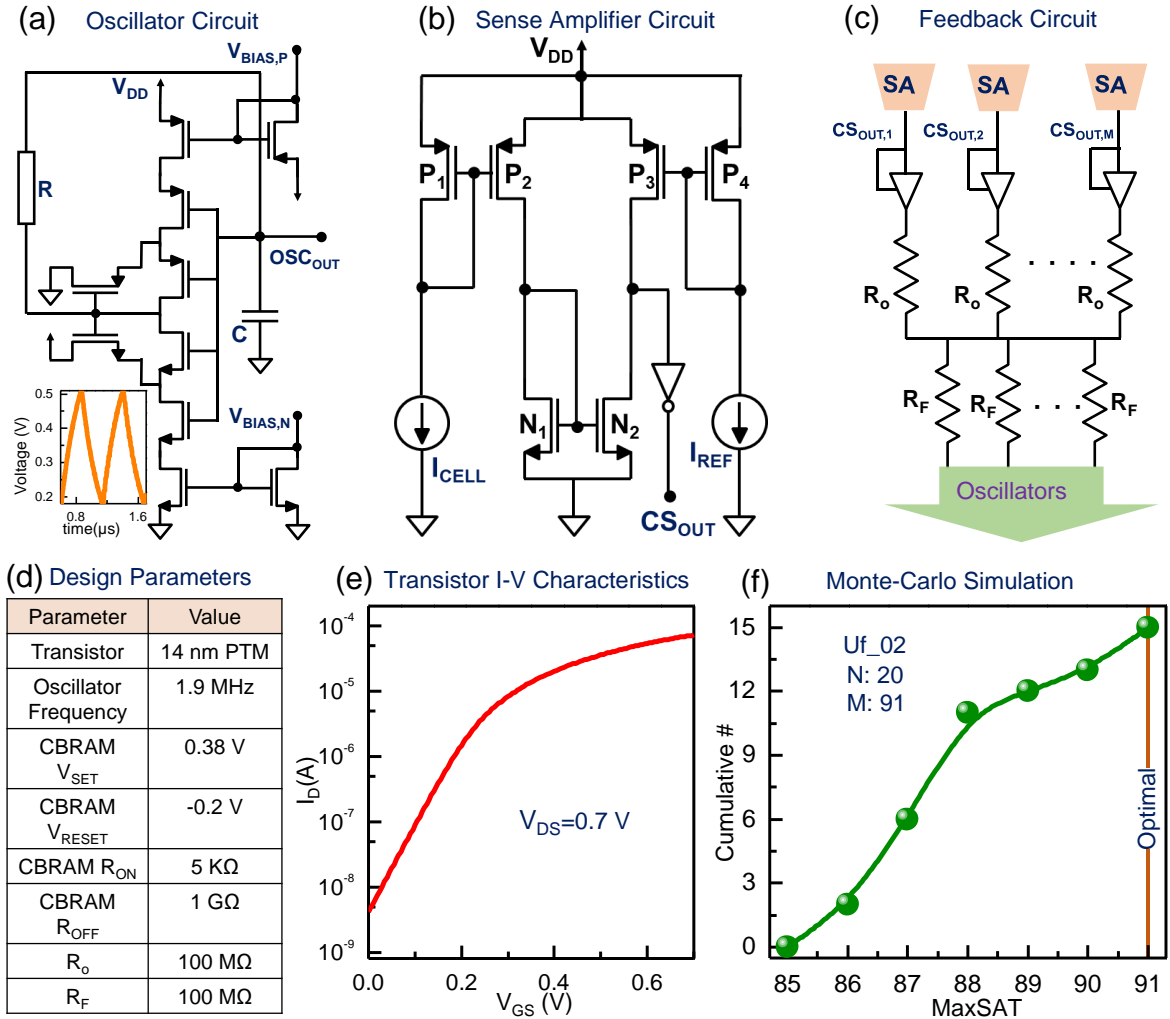

**Fig. S6| CiM implementation. a,** Circuit diagram and the corresponding output waveform (inset) of the Schmitt trigger oscillator used in the CiM-based MaxSAT solver. **b,** Circuit diagram of the sense amplifier (SA) used to read/sense the columns of the NVM array. **c,** Circuit diagram of the adder and the feedback network that enable addition of the output currents of the sense amplifiers, and subsequently, used as feedback to the oscillators. **d,** Table showing the key design parameters of the proposed CiM MaxSAT solver. **e,** I-V characteristics of the 14-nm PTM transistor model for the FET used in the CiM circuit and in the oscillators. **f,** Cumulative distribution of the MaxSAT solutions obtained using Monte-Carlo simulation (15 runs) for a representative MaxSAT instance (uf_02, N=20, M=91).

Here, we discuss the details of the NVM-based implementation for the oscillator-based MaxSAT solver. Figure S6a shows the circuit diagram of the Schmitt trigger oscillator, and the corresponding time domain waveform (inset). Fig. S6b shows the current mirror-based design of the sense amplifier (SA)[3] which is used to evaluate the clause. Fig. S6c shows the resistor-based implementation of the feedback circuit that is used to summate the output (currents) of the SAs, which are subsequently, used as feedback to the oscillators. Key design parameters of the CiM implementation including the (experimentally observed[4]) parameters of the non-volatile RRAM (Cu/HfO$_2$ RRAM) used in the 1T1R cell are shown in Fig. S6d. The I-V characteristics of the 14 nm PTM FET model[5] used to implement the CIM module and the oscillators is shown in Fig. S6e.

To analyze the effect of variation, we perform Monte-Carlo simulations. Fig. S6f shows the cumulative distribution of the measured MaxSAT solution for a representative SAT instance (Uf_20 with N=20, M=91, taken from the SATLIB benchmarking database[6]) over 15 runs. A variation of 10% in the values of the (i) resistance and capacitance of the feedback circuit of the oscillator; (ii) high and low state resistance as well as set and reset voltages of the CBRAM; and (iii) resistors in the feedback circuit (analog adder) of the SAT solver, are considered. It can be observed that the oscillators still produce high-quality solutions in the presence of variations.

## S7. Programing a clause and its evaluation with the memory module

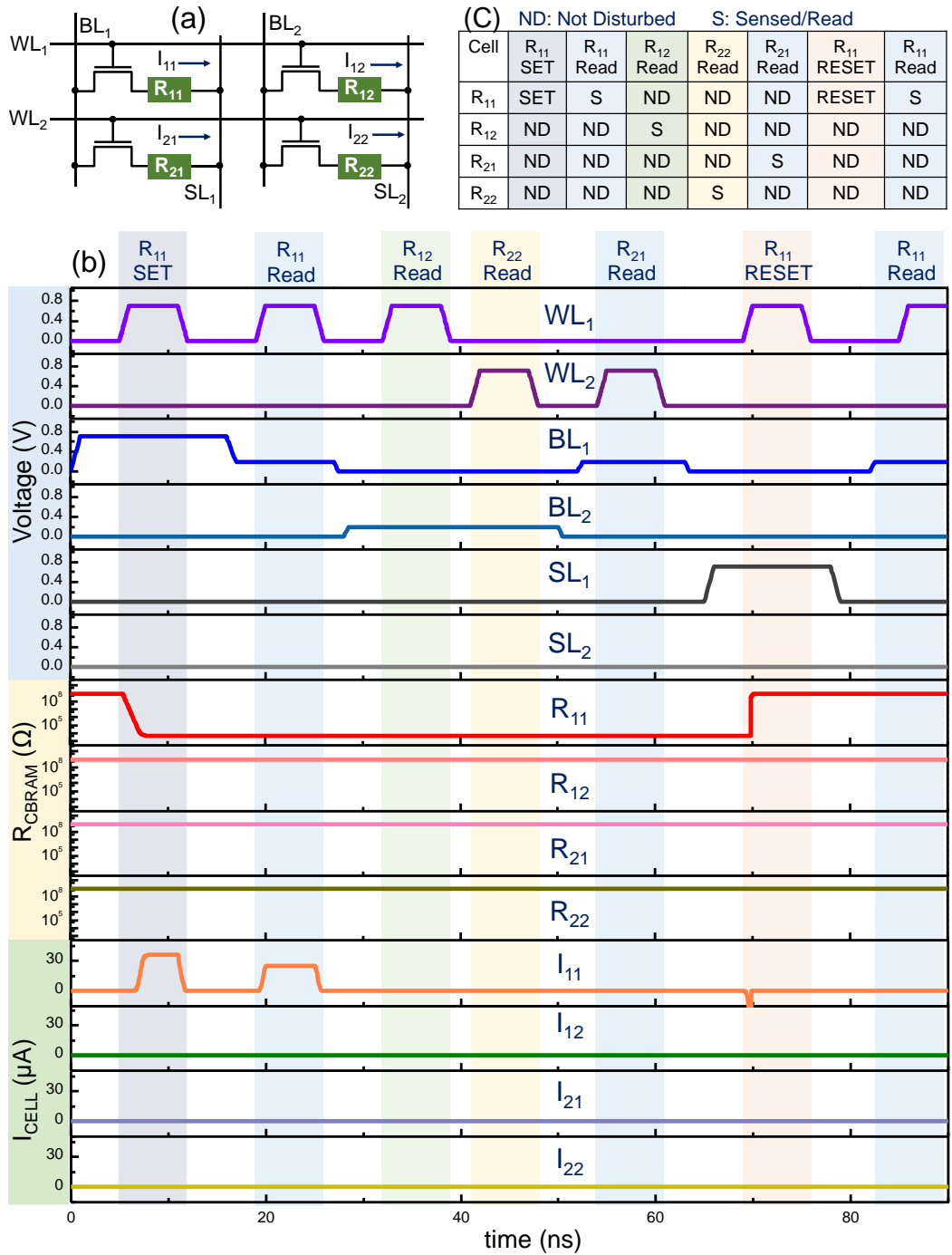

**Fig. S7| Programing a clause and its evaluation in memory. a**, Schematic circuit diagram of a 2x2 CiM module. **b**, Voltage and current waveforms along with corresponding CBRAM resistances illustrating the writing and reading (sensing) scheme; the system can read and write a cell without disturbing the others. **c**, Table summarizing the statuses of each of the cells during the operations shown in the timing diagram in **b**.

Here, we demonstrate the ability to program a CBRAM cell within the memory to represent a variable (during the programming phase) as well as sense the state (during the evaluation phase) without disturbing the other cells in the array. This is illustrated using a 2x2 NVM array (Fig. S7a). Fig. S7b shows the corresponding timing diagram illustrating the ability to program the CBRAM in the cell (here, $R_{11}$) without disturbing the other cells. Additionally, it can be observed that all the four cells can be sensed without disturbing the other cells in the array. The results are summarized in Fig. S7c.

**References**

1. Eleventh Max-SAT Evaluation, retrieved on Feb. 9, 2021.

   At http://maxsat.ia.udl.cat/benchmarks/

2. Max-SAT Benchmark, retrieved on Feb. 9, 2021.

   At http://maxsat.ia.udl.cat/detailed/incomplete-ms-random-table.html

3. Chang, M. F., Shen, S. J., Liu, C. C., Wu, C. W., Lin, Y. F., King, Y. C. *et al.* An offset-tolerant fast-random-read current-sampling-based sense amplifier for small-cell-current nonvolatile memory. *IEEE Journal of Solid-State Circuits* **48**, 864-877 (2013).

4. Shukla, N., Ghosh, R.K., Grisafe, B. & Datta, S. Fundamental mechanism behind volatile and non-volatile switching in metallic conducting bridge RAM. In 2017 IEEE International Electron Devices Meeting (IEDM), 4.3.1-4.3.4, doi: 10.1109/IEDM.2017.8268325 (IEEE, 2017).

5. Latest Models, retrieved on March 5, 2021. At http://ptm.asu.edu

6. SATLIB - Benchmark Problems, retrieved on April 1, 2021.

   At https://www.cs.ubc.ca/~hoos/SATLIB/benchm.html